\documentclass[11pt]{article}

\usepackage[T1]{fontenc}
\usepackage[utf8]{inputenc}

\usepackage{amsmath,amssymb,amsthm,mathtools}
\usepackage{bm}
\usepackage[table]{xcolor}

\usepackage[a4paper,margin=1in]{geometry} 
\usepackage{xcolor}

\usepackage{array}
\usepackage{booktabs}
\usepackage[T1]{fontenc}
\usepackage{amsmath}
\usepackage{tikz}
\usepackage{pgfplots}
\pgfplotsset{compat=1.18}
\usepackage{graphicx}
\usepackage{caption}
\usepackage{subcaption}
\usepackage[table]{xcolor}
\usepackage{tabularx}
\usepackage{array}
\definecolor{lightgray}{gray}{0.9}
\newcolumntype{Y}{>{\raggedright\arraybackslash}X}

\usepackage[colorlinks=true,linkcolor=black,citecolor=blue,urlcolor=blue]{hyperref}

\usepackage{fancyhdr}
\usepackage{authblk}

\newtheorem{theorem}{Theorem}[section]

\newtheorem{proposition}[theorem]{Proposition}

\definecolor{lightgray}{gray}{0.9}
\newtheorem{definition}{Definition}[section]

\usepackage{titlesec}
\titleformat{\section}{\large\bfseries}{\thesection}{0.6em}{}
\titleformat{\subsection}{\normalsize\bfseries}{\thesubsection}{0.6em}{}
\titleformat{\subsubsection}{\normalsize\bfseries}{\thesubsubsection}{0.6em}{}

\usepackage[numbers,sort&compress]{natbib}

\title{\bfseries New Projection Operators for Planar Electrodynamics}

\author[a]{Flávio P. Cruz\thanks{\href{mailto:paiva.cruz@unesp.br}{paiva.cruz@unesp.br}}}
\author[b]{José A. Santos\thanks{\href{mailto:jose.amancio@ifsudestemg.edu.br}{jose.amancio@ifsudestemg.edu.br}}}
\author[b]{Victor J. V. Otoya\thanks{\href{mailto:victor.vasquez@ifsudestemg.edu.br}{victor.vasquez@ifsudestemg.edu.br}}}

\affil[a]{Institute of Science and Technology, UNESP -- Sorocaba Campus\\
Avenue Three of March, 511, Sorocaba--SP, 18087--180, Brazil}

\affil[b]{Institute of Education, Science and Technology of Southeast Minas Gerais\\
Luz Interior Street, 360, Juiz de Fora--MG, 36030--713, Brazil}

\date{} 

\usepackage{float}
\usepackage{tikz}
\usetikzlibrary{calc, math, patterns, arrows.meta}

\begin{document}
\maketitle

\vspace{0.5em}

\begin{abstract}
In this article, we provide a new construction of a set of projection operators and apply it to obtain the propagator of three-dimensional models of electrodynamics; in particular, we consider the Maxwell-Lee-Wick-Chern-Simons, Maxwell-Deser-Jackiw electrodynamics and nonlocal infinite derivative electrodynamics. In order to prove the efficacy of these operators, we perform a causality and unitarity analysis to investigate the presence of ghosts and tachyons.
\end{abstract}



\section{Introduction}\label{sec1}

In Quantum Field Theory (QFT) the calculation of the propagator is essential for the perturbative development in obtaining cross sections \cite{zuber,schwartz,peskin}, as well as for elucidating the nature of the excitations that arise in the interaction between particles and fields \cite{accioly,baeta,boldo}. For this reason, it becomes important to know different methods for calculating these propagators.

It is well known that the pole structure of the propagator encodes the physical dispersion relations of the excitations, since the on-shell energies follow from the zeros of the corresponding denominators in momentum space. Likewise, the analysis of the residues at these poles-in particular, the sign of the residue in the saturated amplitude-provides a direct criterion for tree-level unitarity and is closely tied to the causal properties of the propagating modes \cite{accioly,baeta,boldo}.

In this work, we present an alternative method to obtain the propagator of three-dimensional models. The method is based on a new construction of genuine
projection operators associated with a direct sum
decomposition of subspaces. As initial applications,
we consider Maxwell--Lee--Wick--Chern--Simons
electrodynamics
\cite{borges2022mcs,neves2026lwcs,borges2022lw},
obtained by combining the Lee--Wick higher-derivative
sector
\cite{accioly,leewick,grinstein2008,carone2009,
borges2022mcs,borges2022lw,neves2026lwcs}
with the topological Chern--Simons term introduced
in Ref.~\cite{djackiw}. Such theories have attracted
considerable attention as higher-derivative
extensions capable of improving the ultraviolet
behavior of gauge models while preserving a
consistent particle interpretation through
suitable prescriptions
\cite{leewick,grinstein2008}.
We also consider Maxwell--Deser--Jackiw
electrodynamics \cite{deserjackiw}, which has
been investigated in the context of
higher-derivative planar electrodynamics,
causality and unitarity analysis, and related
gauge-theory extensions
\cite{avila2020,borges2022mcs,lisboasantos2023}.

An important extension of the present work is the
application of the proposed projection-operator
formalism to nonlocal Maxwell--Chern--Simons
electrodynamics. Nonlocal modifications of gauge
theories have been investigated in the context
of effective planar electrodynamics
\cite{Gorbar2023}, while nonpolynomial operator
form factors have also been explored in
higher-derivative field theories
\cite{ModestoRachwal2014}.
Motivated by these developments, we consider
a nonlocal Maxwell--Chern--Simons model
in which the Maxwell and Chern--Simons
sectors are generalized through functions
of the d'Alembertian operator. This extension
provides an opportunity to investigate the
applicability of the proposed projection-operator
formalism beyond local higher-derivative
theories, particularly in the presence of
nonpolynomial momentum-dependent operators.
Furthermore, it allows us to explore how
nonlocal modifications influence the
propagator structure and the corresponding
causality and tree-level unitarity properties
of planar electrodynamics.

The causality and unitarity analysis developed
in this work is closely related to the algorithm
proposed by Accioly and Dias
\cite{acciolydias2005} for topologically massive
higher-derivative electromagnetic models.
In both approaches, the physical spectrum
is determined from the pole structure of
the propagator, and tree-level unitarity is
investigated through the residues of the
saturated propagator. Consequently, the two
formalisms employ equivalent criteria for
identifying physical modes, ghosts, and
tachyons in the corresponding pole analysis.
The main distinction lies in the algebraic
construction of the propagator. While the
Accioly--Dias approach is formulated directly
in terms of the conventional operator
decomposition, in the present work we first
construct a new set of projection operators
and subsequently express the wave operator
as a linear combination of them. As a
consequence, the propagator can be obtained
by exploiting the algebraic properties of
this new set of projection operators.

We begin by presenting a general three-dimensional theory for a vector field, the problem in Section \ref{sec2}, exemplified by the case of Maxwell-Lee-Wick-Chern-Simons electrodynamics \cite{leewick}, taking as a starting point the Lagrangian and obtaining the corresponding field equations in terms of the wave operator.

In the Section \ref{sec3}, we review some concepts and theorems of Linear Algebra, whose demonstrations are found in the Appendix. These theorems are necessary for the construction of the new set of projection operators. Next, in the Section \ref{sec4}, we apply the method to planar electrodynamics and explicitly construct the corresponding projection operators. Later, in the Section \ref{sec5}, we apply the method to obtain the propagator of Maxwell--Lee--Wick--Chern--Simons and Maxwell--Deser--Jackiw electrodynamics. In the Section \ref{sec6}, we present the causality and unitarity analysis for these models. Finally, in the Section \ref{sec7}, we summarize our main results and discuss possible extensions and future directions of the proposed work.

\section{Problem Definition}\label{sec2}
We start with a Lagrangian in the form
\[
\mathcal{L} = \frac{1}{2} A^\mu \mathcal{O}_{\mu\nu} A^\nu - J_\mu A^\mu,
\]
where the wave operator in the momentum space is
\begin{equation}\label{waveoperatorgeneral}
    \mathcal{O}^\mu_{\ \nu}(p) = \alpha(p)\,\omega^\mu_{\ \nu}(p)  + \beta(p)\, \theta^\mu_{\ \nu}(p)+\gamma(p)\,S^{\mu}_{\ \nu}(p).
\end{equation}
Where,
\[
\omega^{\mu}_{\;\nu} = \frac{p^{\mu} p_{\nu}}{p^2},\quad \theta^{\mu}_{\;\nu}= \delta^{\mu}_{\;\nu} - \omega^{\mu}_{\;\nu} \quad \text{and} \quad S^{\mu}_{\;\nu}
= \, \varepsilon^{\mu}_{\,\,\,\nu \beta} \, p^{\beta}.
\]
With this, the functional generator is
\begin{equation}
Z[J] = \mathcal{N} \int \mathcal{D}A \;
\exp \left\{
i \int d^3x\mathcal{L}
\right\}.
\end{equation}
Completing the square and performing the Gaussian integration, we obtain
\begin{equation}
Z[J] = Z[0] \,
\exp \left\{-
\frac{i}{2}
\int d^3x \, d^3y \;
J_\mu(x)\,
G^{\mu}{}_{\nu}(x-y)\,
J^\nu(y)
\right\}.
\end{equation}
The two point functions is
\begin{equation}
\langle 0 | T A_\mu(x) A_\nu(y) | 0 \rangle
= \frac{1}{i^2}
\frac{\delta^2 Z[J]}{\delta J^\mu(x)\, \delta J^\nu(y)}
\bigg|_{J=0}
= iG_{\mu\nu}(x - y).
\end{equation}
Where,
\begin{equation}\label{greenequation}
\mathcal{O}^{\alpha}{}_{\nu}\,
G^{\nu}{}_{\beta}(x - y)
=\delta^{\alpha}{}_{\beta}\,
\delta^{(3)}(x - y).
\end{equation}

Therefore, the calculation of the Green’s functions is basically reduced to the inversion problem of (\ref{waveoperatorgeneral}), which is precisely the goal of this work. For that, we need to understand some mathematical definitions and a little more about projectors. In order to obtain the propagator, our method need some algebraic definitions that will be used for the inversion, the main ones being some results about Projectors and their relation to direct sum decomposition. The proofs of the theorems stated in this section can be found in the Appendix.

\section{Mathematical Formalism}\label{sec3}

The discussion below is grounded on a general algebraic result, namely the Cayley-Hamilton theorem \cite{lang,lima_algebra,axler2024}. Since the vector space $V$ is finite-dimensional, any linear operator acting on it satisfies its own characteristic polynomial. When such an operator is invertible, this implies that its inverse can be written as a linear combination of the identity and its powers. In the context considered here, this theorem provides the conceptual justification for the closure properties assumed for the operators involved and explains why propagators can be consistently expressed within the same algebraic framework.

Given a linear operator $T : V \to V$, where $V$ is a finite-dimensional vector space. Suppose that $T$ is given by a linear combination:

\begin{equation}\label{tsum}
    T = \sum_{i=1}^n \alpha_i T_i,
\end{equation}
which occurs with wave operators where, as in (\ref{waveoperatorgeneral}), we see that $T^{-1}$, if it exists, will be a linear combination of the identity $I$ and products of the operators $T_i$. We are therefore motivated to determine the products $T_i T_j$.

A particularly interesting case occurs when $T_i T_j$ belong to $\text{span}\{T_1, \ldots, T_n\}$ (the subspace generated by the operators $T_i$), that is, $T_i T_j = \sum_k \beta_{ijk} T_k$. In fact, if $T = \sum_i \alpha_i T_i$ is invertible, then $T^{-1} \in \text{span}\{ I, T_1, \ldots, T_n \}$, that is, it will be a linear combination of $I$ and of the $T_i$. We note that, in this case, the subspace $\text{span}\{ I, T_1, \ldots, T_n \}$ will be a subalgebra, that is, if $T, U \in \text{span}\{ I, T_1, \ldots, T_n \}$ and $T$ is invertible, then $TU, T^{-1} \in \text{span}\{ I, T_1, \ldots, T_n \}$.

\begin{definition}
Let $P : V \to V$ be a linear operator. We say that $P$ is a projector if and only if $P^2 = P$. Operators of this type are called Idempotents.
\end{definition}

An immediate consequence is that every vector in the image $\mathrm{Im} \, P = \{ P v ; v \in V \}$ of a projector $P$ is its own eigenvector of eigenvalue $1$, that is, if $v \in \mathrm{Im} \, P$, then $Pv = v$. On the other hand, every vector in the kernel $\ker P = \{ v \in V ; Pv = 0 \}$ is an eigenvector of eigenvalue $0$.

\begin{theorem}\label{theoremdirectsum}
Let $P : V \to V$ be a projector. Then one has:

\begin{equation}\label{directsum}
    V = \mathrm{Im}\, P \oplus \ker P.
\end{equation}
In other words, a projector $P$ decomposes $V$ as a direct sum of its image and kernel. Moreover, $\dim(V) = \dim(\ker P) + \dim(\mathrm{Im}\, P)$.
\end{theorem}

\begin{theorem}\label{theoremnucimg}
Let $P$ be a projector. Then $I - P$ is also a projector. Moreover,

\begin{equation}\label{nucleoimgprojetores}
    \ker P = \mathrm{Im}(I-P), \quad \mathrm{Im} \, P = \ker(I-P).
\end{equation}

\end{theorem}

\begin{theorem}\label{theoreminvO}
Let $\mathcal{C} = \{P_1, P_2, \ldots, P_n\}$
be a set of projection operators satisfying $\sum_{i=1}^{n} P_i = \mathbf{1} \text{ and }
P_i P_j = \delta_{ij}P_i,$
where $\mathbf{1}$ denotes the identity operator.
Suppose that an operator $O$ can be written as a linear combination of the projectors in $\mathcal{C}$, $O = \sum_{i=1}^{n} \alpha_i P_i,$
with $\alpha_i \neq 0 \, \forall \, i$.
Then $O$ is invertible and its inverse $O^{-1}$ is given by

\begin{equation}\label{invO}
   O^{-1} = \sum_{i=1}^{n} \alpha_i^{-1} P_i. 
\end{equation}

\end{theorem}

\section{Solutions of the Problem}\label{sec4}

Since the wave operator in momentum space is given by (\ref{waveoperatorgeneral}), where $\alpha(p), \beta(p), \gamma(p) \in \mathbb{C}$, our objective is to derive its inverse in a form valid for arbitrary coefficient functions. For this purpose, we shall begin a deeper study of the operators $\omega, \theta, S : \mathbb{C}^3 \to \mathbb{C}^3$, defined by $(\omega x)^{\mu} = \omega^{\mu}_{\;\nu} x^{\nu}$,  $(\theta x)^{\mu} = \theta^{\mu}_{\;\nu} x^{\nu}$ and $(Sx)^{\mu} = S^{\mu}_{\;\nu} x^{\nu}$, where $x = (x^0, x^{1}, x^{2}) \in \mathbb{C}^3$, in order to obtain properties that will facilitate the obtaining of $\mathcal{O}^{-1}$.

\subsection{Operators $\omega$, $\theta$ and $S$}

These operators satisfy the following multiplication table.

\begin{table}[htb]

    \centering 
    \caption{Products between $\omega$, $\theta$ and S.}\label{tabLTB}
    \vspace{3mm} 

    \begin{tabular}{ | l | l | l | l | l | l | l | l | l | p{1cm} |}
        \hline
          & $\omega$ & $\theta$ & \,\,\,\,\,\,S  \\ \hline
        $\omega$ & $\omega$ & 0 & \,\,\,\,\,\,0  \\ \hline
        $\theta$ & 0 & $\theta$ & \,\,\,\,\,\,S  \\ \hline
        S & 0 & S & $- p^2 \theta$ \\ \hline
    \end{tabular}

\end{table}
Furthermore, it should be noted that
$\omega$ and $\theta=I-\omega$ are orthogonal projectors. Thus, from theorems \ref{theoremdirectsum} and \ref{theoremnucimg} we have $\mathrm{Im}\, \omega = \ker \theta$, $\ker \omega = \mathrm{Im}\, \theta$, $\mathbb{C}^3 = \mathrm{Im}\, \omega \oplus \mathrm{Im}\, \theta$ and the following relations are satisfied

\begin{itemize}
    \item $dim(\mathrm{Im}\, \omega) = 1 \text{ and } dim(\ker \omega) =2 \quad \forall \,\, p \neq 0.$
    \item $dim(\mathrm{Im}\, \theta) = 2 \text{ and } dim(\ker \theta) =1.$
\end{itemize}

\begin{proposition}\label{proposition1}

The subspaces $\mathrm{Im}\,\omega$ and $W=\mathrm{Im}\,\theta=\ker \omega$ are invariant under $\omega, \theta$ and $S$.  
Let $S_W = S|_W : W \to W$ and $\mathbf{1}_W = \mathbf{1}_W : W \to W$ be the restrictions of $S$ and of the identity operator $\mathbf{1}$ to the subspace $W$.  
Then the operators $Z_{\pm} : W \to W$ defined by:

\begin{equation}\label{ppm}
    Z_{\pm} = \frac{1}{2}\,\mathbf{1}_W \pm \frac{i}{2\sqrt{p^2}}\, S_W
\end{equation}
are projectors and therefore we have the decompositions
\[
W = \mathrm{Im}\,Z_{\pm} \oplus \ker Z_{\pm}.
\]
Moreover, such decomposition is non-trivial, that is,
\[
\dim(\mathrm{Im}\,Z_{\pm}) = \dim(\ker Z_{\mp}) = 1,
\]
and $\mathrm{Im}\,Z_{\pm}=\ker Z_{\mp}$ are eigenspaces of $S$ with eigenvalues $-i\sqrt{p^2} \text{ and } i\sqrt{p^2},$
respectively.
    
\end{proposition}

\section{Calculation of the Propagators}\label{sec5}

First, let us begin by defining the following projection operators $\{P_1=\omega,P_2=Z_+\theta, P_3=Z_-\theta\}$. Or, we can write them explicitly as

\[
P_1=\omega,
\]
\[
P_2 = \frac{1}{2}\,\theta + \frac{i}{2\sqrt{p^2}}\, S,
\]

\[
P_3 = \frac{1}{2}\,\theta - \frac{i}{2\sqrt{p^2}}\, S.
\]
It should be noted that these operators satisfy the completeness relation

\begin{equation}\label{sumomegazmaiszmenos}
    \sum_i P_i = \mathbf{1}.
\end{equation}
And if we apply these operators to each other, we conclude that the orthogonality relationship is also satisfied
\begin{equation}
    P_i P_j = \delta_{ij} P_i.
\end{equation}

Now, writing the operators $\theta$ and $S$ in terms of the $P_2\text{ and } P_3$, we obtain

\begin{equation}
    \theta = P_2+P_3,
\quad S = -i\sqrt{p^2}(P_2 - P_3).
\end{equation}

\subsection{Propagator for arbitrary coefficients}\label{subsec5.1}

Substituting these relations into equation (\ref{waveoperatorgeneral}), we obtain

\begin{equation}\label{MPs}
    \mathcal{O} = \alpha P_1 + (\beta - i\gamma \sqrt{p^2}) P_2 + (\beta + i\gamma \sqrt{p^2}) P_3,
\end{equation}
Consequently, one can write the general propagator in terms of $P_1,P_2$ and $P_3$,

\begin{equation}\label{propagator1}
    \mathcal{O}^{-1}= \frac{1}{\alpha} P_1 + \frac{1}{\beta - i\gamma \sqrt{p^2}} P_2 + \frac{1}{\beta + i\gamma \sqrt{p^2}} P_3.
\end{equation}
Thus, going back to $\omega,\theta$ and $S$, we obtain

\begin{equation}\label{generalpropagator}
    \mathcal{O}^{-1}
= \frac{1}{\alpha}\,\omega
+ \frac{\beta}{\beta^{2}+\gamma^{2}p^{2}}\,\theta
- \frac{\gamma}{\beta^{2}+\gamma^{2}p^{2}}\,S.
\end{equation}

\subsection{The Maxwell-Lee-Wick-Chern-Simons Propagator}

We start from the Lagrangian of Maxwell-Lee-Wick-Chern-Simons electrodynamics \cite{accioly, neves2026lwcs},

\[
\mathcal{L} = -\frac{1}{4} F_{\mu\nu} F^{\mu\nu} - \frac{\mu}{4}F_{\mu\nu} \Box F^{\mu\nu}
\]
\begin{equation}\label{lagrangian}
- \frac{1}{2\xi} (\partial^\mu A_\mu)^2 + \frac{\kappa}{2} \epsilon^{\mu\nu\rho} A_\mu \partial_\nu A_\rho
- J_\mu A^\mu,
\end{equation}
where $A^\mu$ is the vector potential, 
$F_{\mu\nu} = \partial_\mu A_\nu - \partial_\nu A_\mu$ is the electromagnetic field tensor and $J_\mu$ an external current. 
The gauge-fixing term, with the parameter $\xi$, must be included in order to make the calculation of the propagator possible. In fact, as its known, without this term the wave operator would not be invertible.
In this way, using the coefficients of corresponding wave operator,

\begin{equation}\label{abgMLWCS}
    \alpha = -\frac{p^2}{\xi}, \quad \beta=-p^2+\mu\,p^4 \quad \text{and} \quad \gamma = i\kappa,
\end{equation}
we have the propagator for this theory

\[
    \mathcal{O}^{-1}= -\frac{\xi}{p^2} P_1 + \frac{1}{-p^2+\mu\,p^4  + \kappa \sqrt{p^2}} P_2 
\]

\begin{equation}\label{propagatorps1}
+ \frac{1}{-p^2+\mu\,p^4  - \kappa \sqrt{p^2}} P_3.
\end{equation}
Thus, at the basis $\{\omega, \theta, S \}$, we have

\[
\mathcal{O}^{-1}_{\mu\nu}
= -\frac{p^2\left(1-\mu\,p^2\right)}
{p^2\left[p^2\left(1-\mu\,p^2\right)^2-\kappa^2\right]}
\,\theta_{\mu\nu}
\]
\begin{equation}\label{propagator2}
 -
\frac{i\kappa}
{p^2\left[p^2\left(1-\mu\,p^2\right)^2-\kappa^2\right]}
\,S_{\mu\nu}
-
\frac{\xi}{p^2}\,\omega_{\mu\nu}.
\end{equation}

\subsection{The Maxwell-Deser-Jackiw Propagator}

Another interesting model can be built up from the Maxwell system by adding the higher derivative Chern-Simons extension proposed by \cite{deserjackiw, borges2022mcs}

\[
    \mathcal{L}=-\frac{1}{4} F_{\mu\nu} F^{\mu\nu}+\frac{1}{2m}\, \varepsilon^{\mu\nu\lambda} (\Box A_\mu)(\partial_\nu A_\lambda)
\]
\begin{equation}\label{lagrangiandj}
-\frac{1}{2\xi} (\partial_\mu A^\mu)^2
- J^\mu A_\mu .
\end{equation}
The corresponding wave operator has the following coefficients

\begin{equation}\label{abgMDJ}
    \alpha = -\frac{p^2}{\xi},\quad \beta = -p^2 \quad \text{and} \quad \gamma = i\frac{p^2}{m}.
\end{equation}
Thus, the propagator of this theory is

\[
    \mathcal{O}^{-1}= -\frac{\xi}{p^2} P_1 - \frac{1}{p^2 - \tfrac{p^2 \sqrt{p^2}}{m}} P_2 
\]

\begin{equation}\label{propagatorps1_alt}
- \frac{1}{p^2 + \tfrac{p^2 \sqrt{p^2}}{m}} P_3.
\end{equation}
At the basis $\{\omega, \theta, S \}$, with $\gamma = ip^2/m$, we have

\begin{equation}\label{propagator3}
\mathcal{O}^{-1}_{\mu\nu}(p)
= -\,\frac{1}{p^{2}\!\left(1 - \frac{p^{2}}{m^{2}}\right)}
\left( \theta_{\mu\nu} + \frac{i}{m}\,S_{\mu\nu} \right)
- \frac{\xi}{p^{2}}\,\omega_{\mu\nu}.
\end{equation}

\subsection{The Nonlocal Maxwell--Chern--Simons
Propagator}\label{subsec:5.4}

The nonlocal Maxwell--Chern--Simons model is described by the following Lagrangian density:
\begin{equation}
\begin{aligned}
\mathcal{L}
=
-\frac{1}{4}
F_{\mu\nu}
e^{-\Box/M^{2}}
F^{\mu\nu}
+
\frac{\kappa}{2}
\epsilon^{\mu\nu\rho}
A_{\mu}
f(\Box)
\partial_{\nu}A_{\rho}
- \frac{1}{2\xi} (\partial^\mu A_\mu)^2
-
J_{\mu}A^{\mu},
\end{aligned}
\label{eq:nonlocal_lagrangian}
\end{equation}
where $M$ denotes the characteristic nonlocality
scale. The function $f(\Box)$ is assumed to admit the
power-series representation
\begin{equation}
f(\Box)
=
\sum_{n=0}^{\infty}
c_n
\left(
\frac{\Box}{M^2}
\right)^n,
\label{eq:nonlocal_form_factor}
\end{equation}
while the exponential function is defined by
\begin{equation}
e^{-\Box/M^2}
=
\sum_{n=0}^{\infty}
\frac{(-1)^n}{n!}
\left(
\frac{\Box}{M^2}
\right)^n.
\label{eq:nonlocal_exponential}
\end{equation}
The equivalent wave operator has the following coefficients
\begin{equation}\label{abgn}
    \alpha = -\frac{p^2}{\xi},\quad
    \beta = -p^2 e^{p^2/M^2}
    \quad \text{and} \quad
    \gamma = i\kappa f\left(-p^2\right).
\end{equation}
Thus, the propagator of this theory is
\begin{equation}\label{propagatorps1_nl}
\begin{aligned}
\mathcal{O}^{-1}(p)
={}&
-\frac{\xi}{p^2}P_1
\\[2mm]
&+
\frac{P_2}{
-p^2e^{p^2/M^2}
+
\kappa f(-p^2)\sqrt{p^2}
}
\\[2mm]
&+
\frac{P_3}{
-p^2e^{p^2/M^2}
-
\kappa f(-p^2)\sqrt{p^2}
}.
\end{aligned}
\end{equation}
This expression constitutes the propagator
of the nonlocal Maxwell--Chern--Simons
model in the new projection-operator basis. It is worth emphasizing that the
nonlocal form factors modify the scalar
coefficients of the wave operator without
changing the algebraic properties of the
projectors. Consequently, the inversion procedure
remains identical to that employed for
the local higher-derivative models. For completeness, we now express the
propagator in the conventional
$\{\omega,\theta,S\}$ basis. Using the general result
\begin{equation}
\mathcal{O}^{-1}
=
\frac{1}{\alpha}\omega
+
\frac{\beta}{
\beta^2+\gamma^2p^2
}\theta
-
\frac{\gamma}{
\beta^2+\gamma^2p^2
}S,
\label{eq:nonlocal_general_inverse}
\end{equation}
we first calculate the common
transverse denominator,
\begin{equation}
\begin{aligned}
D(p)
&=
\beta(p)^2+\gamma(p)^2p^2
\\
&=
p^4e^{2p^2/M^2}
-
\kappa^2f(-p^2)^2p^2
\\
&=
p^2
\left[
p^2e^{2p^2/M^2}
-
\kappa^2f(-p^2)^2
\right].
\end{aligned}
\label{eq:nonlocal_denominator}
\end{equation}
Hence, the propagator in the conventional
operator basis takes the form
\begin{equation}
\begin{aligned}
\mathcal{O}^{-1}_{\mu\nu}(p)
={}&
-
\frac{
e^{p^2/M^2}
}{
p^2e^{2p^2/M^2}
-
\kappa^2f(-p^2)^2
}
\theta_{\mu\nu}(p)
\\[2mm]
&-
\frac{
i\kappa f(-p^2)
}{
p^2
\left[
p^2e^{2p^2/M^2}
-
\kappa^2f(-p^2)^2
\right]
}
S_{\mu\nu}(p)
\\[2mm]
&-
\frac{\xi}{p^2}
\omega_{\mu\nu}(p).
\end{aligned}
\label{eq:nonlocal_final_propagator}
\end{equation}
As a consistency check, we consider the local limit
$M\to\infty$. For fixed momentum and $c_0=1$,
the nonlocal functions approach unity, and the
propagator reduces to that of the conventional
Maxwell--Chern--Simons theory, in agreement with
the results obtained in Section \ref{subsec5.1}.

\section{Causality and Unitarity Analysis in the General-Coefficient Framework}
\label{sec6}

In this section we present the
analysis of causality and tree–level unitarity for a class of
planar gauge theories whose wave operator in momentum space can be
written as (\ref{waveoperatorgeneral}). The analysis presented here follows a standard framework for testing causality and tree-level unitarity in field theories, closely related to
the procedure proposed by Accioly and Dias~\cite{acciolydias2005} for
topologically massive higher-derivative models. Unitarity is investigated through the saturation of the propagator with conserved external currents and the inspection of the pole structure of the resulting current--current amplitude, requiring the positivity of the residues in order to exclude ghost excitations, as discussed in higher-derivative gravity models \cite{azeredo}. 
On the other hand, causality is implemented through the requirement that the group velocity of physical excitations does not exceed the speed of light, $v_g \leq 1$, which provides a criterion applicable to theories described by the
general-coefficient wave operator based on the dispersion relation of the propagating modes \cite{shabad2011,shabad2009}. 
This condition ensures that the propagation remains subluminal and is consistent with the general principles governing the analytic structure of relativistic field theories.

\subsection{Analysis on the basis $\{P_1,P_2,P_3\}$}

Let $J_\mu(p)$ be an external conserved current, $p^\mu J_\mu(p)=0$, the saturated amplitude is then

\begin{equation}
\mathcal A(p) =
J^{*\mu}(p)\,\mathcal{O}^{-1}_{\mu\nu}(p)\,J^\nu(p).
\end{equation}
Using the form (\ref{propagator1}), we restrict the analysis to the physical time-like region
$p^2>0$, which corresponds to massive propagating modes.
Space-like momenta ($p^2<0$) do not lead to physical on-shell excitations and
therefore do not contribute to the unitarity analysis at tree level.
In this regime we obtain
\begin{equation}
\mathcal A(p)
=
-\frac{\|J^{(+)}\|^2}{\beta(p)-i\gamma(p)\sqrt{p^2}}
-
\frac{\|J^{(-)}\|^2}{\beta(p)+i\gamma(p)\sqrt{p^2}},
\label{eq:A_general_P}
\end{equation}
where $J^{(+)}=P_2J$, $J^{(-)}=P_3J$ and the norms
$\|J^{(\pm)}\|^2=-\, (J^{(\pm)})^{*\mu}\theta_{\mu\nu}J^{(\pm)\nu}\ge 0$ are taken in the
physical transverse subspace. Thus, we will call the coefficient of each term in (\ref{eq:A_general_P}) as follows

\[
F_+(p)=\frac{1}{\beta(p)-i\gamma(p)\sqrt{p^2}},
\]

\begin{equation}
\qquad
F_-(p)=\frac{1}{\beta(p)+i\gamma(p)\sqrt{p^2}}.
\label{eq:scalar-props}
\end{equation}

\subsubsection{Pole structure}
\label{subsec:poles-general}

The poles are obtained from the denominators of the saturated amplitude~\eqref{eq:A_general_P}. Since the longitudinal contribution vanishes when
the propagator is saturated with a conserved current, only the
transverse sector contributes to the physical amplitude. Multiplying the conjugate factors, we obtain
\begin{equation}
 D(p) \equiv \bigl[\beta(p)-i\gamma(p)\sqrt{p^2}\bigr]
\bigl[\beta(p)+i\gamma(p)\sqrt{p^2}\bigr]
= \beta(p)^2+\gamma(p)^2 p^2.
\end{equation}
Therefore, the poles associated with the vanishing of $D(p)$ are located at
momentum
$p^\mu=p_*^\mu$ satisfying

\begin{equation}
D(p_*)=0,
\label{eq:pole-condition}
\end{equation}
i.e., at the solutions of

\begin{equation}
\beta(p_*)^2 + \gamma(p_*)^2\,p_*^2 = 0.
\end{equation}
For each fixed spatial momentum $\vec p$ these correspond to energies
$p^0=E_*(\vec p)$ given by the solutions of

\begin{equation}
D(p^0,\vec p)=0.
\end{equation}

\paragraph{Real poles}
If $E_*(\vec p)$ is real, we obtain a propagating mode with dispersion
relation determined implicitly by $D(E_*(\vec{p}),\vec p)=0$. For such modes,
causality at the microscopic level requires that the group velocity

\begin{equation}
v_g = \left|\frac{\partial E_*(\vec{p})}{\partial \vec p}\right|
\leq 1,
\label{eq:vg-condition}
\end{equation}
which can be translated into inequalities involving the derivatives of
$D$ with respect to $p^0$ and $\vec p$ using

\begin{equation}
\frac{\partial D}{\partial p^0}\Big|_{p_*}\,dE_*(\vec{p})
+
\frac{\partial D}{\partial \vec p}\Big|_{p_*}\!\cdot d\vec p = 0
\quad\Rightarrow\quad
\frac{\partial E_*(\vec{p})}{\partial \vec p}
=
-\,\frac{\partial D/\partial \vec p}{\partial D/\partial p^0}\Bigg|_{p_*}.
\end{equation}

\paragraph{Complex poles}
If some solutions $E_*(\vec p)$ are complex, they occur in conjugate
pairs by hermiticity of the underlying Lagrangian. Writing
$E_*(\vec p)=\Omega_n(\vec p)-i\Gamma_n(\vec p)$ with
$\Gamma_n(\vec p)>0$, the corresponding Fourier modes behave as
$\exp[-i\Omega_n t]\,\exp[-\Gamma_n t]$ and can be interpreted this later on.

\subsubsection{Residues and tree–level unitarity}
\label{subsec:residues-general}

To analyse unitarity we focus on the behaviour of the propagator near
its poles. For a fixed $\vec p$, consider a simple pole at
$p^\mu=p_*^\mu$ such that

\begin{equation}
D(p_*)=0,\qquad
\left.\frac{\partial D}{\partial p^0}\right|_{p_*}\neq 0.
\end{equation}
Expanding $D$ around $p^0=E_*(\vec p)$ we have
\begin{equation}
D(p^0,\vec p) \simeq
\left.\frac{\partial D}{\partial p^0}\right|_{p_*}
\bigl(p^0-E_*(\vec p)\bigr).
\end{equation}
Using (\ref{eq:scalar-props}), the scalar propagators near the pole behave as

\begin{equation}
F_\pm(p^0,\vec p)
\simeq
\frac{\beta(p_*)\pm i\gamma(p_*)\sqrt{p_*^2}}%
     {\left.\dfrac{\partial D}{\partial p^0}\right|_{p_*}}\,
\frac{1}{p^0-E_*(\vec{p})}.
\end{equation}
Therefore, the residues of the two contributions $\mathcal A_\pm$ to the saturated
amplitude at $p^0=E_*(\vec{p})$ are

\begin{equation}
\text{Res}\,\mathcal A_\pm\Big|_{p^0=E_*(\vec{p})}
=-
\|J^{(\pm)}\|^2\,
\frac{\beta(p_*)\pm i\gamma(p_*)\sqrt{p_*^2}}%
     {\left.\dfrac{\partial D}{\partial p^0}\right|_{p_*}},
\label{eq:ResA-general}
\end{equation}
with $\|J^{(\pm)}\|^2\ge 0$. At this point it is convenient to adopt a more compact formulation.
We reorganize the saturated amplitude so that the contributions associated
with $D_+(p)$ and $D_-(p)$ appear explicitly,

\begin{equation}
\mathcal A(p) = -\frac{D_{-}(p)\,\|J^{(+)}\|^{2} + D_{+}(p)\,\|J^{(-)}\|^{2}}
     { D(p)},
\end{equation}
where $D_{\pm}(p) = \beta(p)\mp i\gamma(p)\sqrt{p^{2}}$.
Therefore, the residue of the saturated amplitude at $p^{0}=E_*(\vec{p})$ follows directly from the expansion of $D(p)$,
\begin{equation}
\text{Res}\,\mathcal A\Big|_{p^0=E_*(\vec{p})}
=-\frac{D_{-}(p_{*})\,\|J^{(+)}\|^{2} + D_{+}(p_{*})\,\|J^{(-)}\|^{2}}
     {\dfrac{\partial D}{\partial p^0}(p_{*})}.
\end{equation}

Since the pole condition implies either $D_{+}(p_{*})=0$ or $D_{-}(p_{*})=0$, the numerator simplifies accordingly. Indeed, 
\begin{eqnarray*}
D_{+}(p_{*})=0 \quad &\Rightarrow& \quad D_{-}(p_{*}) = 2\beta(p_{*}) = 2i\gamma(p_{*})\sqrt{p_{*}^{2}},\\
D_{-}(p_{*})=0 \quad &\Rightarrow& \quad D_{+}(p_{*}) = 2\beta(p_{*}) = -2i\gamma(p_{*})\sqrt{p_{*}^{2}}.
\end{eqnarray*}

Thus, we obtain the following expressions for the residue
\begin{equation}\label{resAcompact}
    \text{Res}\,\mathcal A\Big|_{p^0=E_*(\vec{p})} = 
    -\frac{2\,\beta(p_{*})} {\partial_{0}{D}(p_{*})} \, \|J^{(\sigma)}\|^{2} = 
    -\frac{\|J^{(\sigma)}\|^{2}} {\partial_{0}{D_{\sigma}}(p_{*})}, 
    \quad \sigma =
    \begin{cases}
        +,& D_{+}(p_{*}) = 0,\\
        -,& D_{-}(p_{*}) = 0.
    \end{cases} 
\end{equation}
Since $\|J^{(\pm)}\|^2\ge 0$, this means, according to the first expression, that the residue sign is determined by the sign of $-\beta(p_{*})/\partial_{0}{D}(p_{*})$. For real poles, the saturated amplitude is real. In the models considered here, where $\gamma(p)$ is purely imaginary, each contribution is real separately, and no complex-conjugate pairing between the residues is required. Tree–level unitarity requires that the terms of~\eqref{resAcompact} be non-negative.

\subsubsection{Equivalent analysis in the basis $\{\omega,\theta,S\}$}
\label{subsec:ots-analysis}

For completeness we show how the same conclusions follow directly
from the form~\eqref{generalpropagator} of the propagator.
Saturating with a conserved current $J_\mu$ and using
$\omega J = 0$ and $J=\theta J$, we obtain

\begin{equation}
\mathcal A(p) =
J^*\!\cdot \mathcal{O}^{-1}\! J
=
\frac{\beta(p)}{D(p)}\,J^*\!\cdot J
-
\frac{\gamma(p)}{D(p)}\,J^*\!\cdot S\! J.
\label{eq:A-ots}
\end{equation}
In the transverse
subspace we can again decompose $J$ into eigenvectors

\begin{equation}
J = J^{(+)}+J^{(-)},\qquad
S J^{(\pm)} = \mp i\sqrt{p^2}\,J^{(\pm)}.
\end{equation}
It follows that

\begin{equation}
J^*\!\cdot J = -\|J^{(+)}\|^2-\|J^{(-)}\|^2,
\end{equation}
and

\[
J^*\!\cdot S\! J
=
(J^{(+)})^*\!\cdot S J^{(+)}
+
(J^{(-)})^*\!\cdot S J^{(-)}
=
\]

\begin{equation}
=
i\sqrt{p^2}\,\|J^{(+)}\|^2
-
i\sqrt{p^2}\,\|J^{(-)}\|^2.
\end{equation}
Substituting into ~\eqref{eq:A-ots} we obtain

\[
\mathcal A(p)
=-\frac{1}{D(p)}
\Bigl[
   \bigl(\beta(p)+i\gamma(p)\sqrt{p^2}\bigr)\,\|J^{(+)}\|^2\Bigr]
\]

\begin{equation}
-
\frac{1}{D(p)}
\Bigl[
\bigl(\beta(p)-i\gamma(p)\sqrt{p^2}\bigr)\,\|J^{(-)}\|^2
\Bigr],
\end{equation}
we recover exactly (\ref{eq:A_general_P}). Thus the
decomposition and the analysis of poles, residues, unitarity and
causality are entirely equivalent in the $\{P_1,P_2,P_3\}$ and
$\{\omega,\theta,S\}$ basis. The projector formalism therefore
provides a compact and geometrically transparent framework for
studying higher–derivative planar gauge models at the level of their
propagators.

\subsection{Analysis for the Maxwell--Lee--Wick--Chern--Simons model}

Using (\ref{abgMLWCS}), the physical poles of the transverse sector are therefore the solutions of
\begin{equation}
p^2=0,\qquad
p^2\!\left(1-\mu p^2\right)^2-\kappa^2=0.
\label{eq:MLWCS-pole-eq}
\end{equation}
The pole at $p^2=0 \Rightarrow p^0=\pm \|\vec{p}\|$ is purely a gauge pole and disappears when the
propagator is saturated with a conserved current, so that the
physical structure is determined by the cubic term in $p^2$. In this sense, it is important to note that at $p^{2}=0$ the standard decompositions based on $\omega_{\mu\nu}$ and on the projector basis associated with the transverse subspace are no longer uniformly well defined. Consequently, the analysis of the massless pole must be carried out either by a careful limiting procedure or by adopting a light-like parametrization of the external sources. Although the propagator exhibits explicit $1/p^{2}$ factors, including in terms proportional to $S_{\mu\nu}$, saturation with conserved currents removes the pure gauge sector ($\omega_{\mu\nu}$), while the contributions involving $S_{\mu\nu}$ are suppressed, since $S_{\mu\nu}\sim\epsilon_{\mu\nu\rho}p^{\rho}$. The remaining $1/p^{2}$ behavior in the saturated amplitude $\mathcal A(p)$ should therefore be interpreted as a long-range interaction rather than as a genuine propagating massless degree of freedom, and the appropriate criterion for massless unitarity is the evaluation of the sources together with the gauge-induced constraints, as in the approach of \cite{accioly}. For each fixed $\vec{p}$, defining $x=\mu((p^0)^2-\vec p^{\,2})$ and $\beta\equiv\kappa^2\mu$, ~\eqref{eq:MLWCS-pole-eq}
becomes

\begin{equation}
x\left(1-x\right)^2-\beta=0.
\end{equation}
This cubic equation has three real roots for $0\le\beta<4/27$,
two degenerate real roots for $\beta=4/27$ and only one real root
for $\beta>4/27$. In terms of the original parameters:
\begin{itemize}
\item $\displaystyle \kappa^2\mu <\frac{4}{27}$:
 three real massive modes (two of them with negative residue, as we will
 see below);
\item $\displaystyle \kappa^2\mu=\frac{4}{27}$:
 critical case, with two degenerate modes;
\item $\displaystyle \kappa^2\mu>\frac{4}{27}$:
 one real massive mode and one pair of complex-conjugate poles
 (Lee--Wick pair);
\item $\displaystyle \kappa^2\mu<0$:
a single real massive mode in the physical region $p^2>0$.
\end{itemize}
Each real root $x_i$ define $p_i^2=x_i/\mu$. Therefore the dispersion relationship is $p^0=\pm \sqrt{\vec p^{\,2}+p_i^2}$.

\subsubsection{Saturated amplitude and residues}

For a conserved current, the longitudinal sector ($P_1$) does not
contribute to the amplitude. Using 
~\eqref{eq:A_general_P} with the coefficients
above, the saturated amplitude is

\[
\mathcal A(p) = \frac{\|J^{(+)}\|^2}{(p^0)^2 - \vec{p}^{\,2} 
- \mu\big((p^0)^2 - \vec{p}^{\,2}\big)^2
- \kappa\,\sqrt{(p^0)^2 - \vec{p}^{\,2}}}
\]
\begin{equation}
    + \frac{\|J^{(-)}\|^2}{(p^0)^2 - \vec{p}^{\,2} 
- \mu\big((p^0)^2 - \vec{p}^{\,2}\big)^2
+ \kappa\,\sqrt{(p^0)^2 - \vec{p}^{\,2}}}.
\end{equation}
Defining the product of denominators $D(p^0,\vec{p})=\left[(p^0)^2-\vec p^{\,2}\right] G(p^2)$, the residue of the saturated amplitude at $p^0=E_*(\vec{p})$ is

\begin{equation}
\operatorname*{Res}_{\,p^{0}=E_*(\vec p)} \mathcal A(p^{0},\vec p)
=
\frac{\left(1-\mu \lambda\right)}{G'(\lambda)E_*(\vec{p})}\|J^{(\sigma)}\|^2,
\qquad
\sigma=
\begin{cases}
+,& \kappa\left(1-\mu \lambda\right)>0,\\
-,& \kappa\left(1-\mu \lambda\right)<0.
\end{cases}
\label{eq:residue_general_kappa_sign}
\end{equation}

Where $
G'(\lambda) = \left(1-\mu \lambda\right)\left(1-3\mu \lambda\right).$
Since $E_*(\vec{p})>0$, we see that the sign of the residue is controlled by

\begin{equation}\label{sngres}
\text{sgn}\,\bigl(\text{Res}\,\mathcal A\bigr)
= \text{sgn}\!\left[
 \,\frac{1}{1-3\mu \lambda}
\right].
\end{equation}

Now, consider $D_\pm(p)=-p^2(1-\mu p^2)\pm \kappa\sqrt{p^2},$
in the physical region $p^2>0$. Introducing $t=\sqrt{p^2}>0$, we rewrite
\begin{equation}
D_\pm(p)=-t^2(1-\mu t^2)\pm \kappa t
= t\bigl[\mu t^3-t\pm \kappa\bigr].
\label{eq:Dpm-t}
\end{equation}
Hence, besides the massless factor $t=0$, the massive poles in $p^2>0$
are determined by the cubic equations
\begin{equation}
f_\pm(t):=\mu t^3-t\pm\kappa=0,
\qquad t>0,
\label{eq:fpm}
\end{equation}

whose dependence on the parameters $\mu$ and $\kappa$ is analysed below. Considering the auxiliary function $f_{0}(t) = t\,(\mu \, t^{2} - 1)$, and the result $f'_{\pm}(t) = f'_{0}(t) = 3\mu  t^{2} - 1,$ we obtain the following results that for the corresponding negative $\kappa$ values can be obtained interchanging $f_{+}\leftrightarrow f_{-}$. The Figure \ref{fig:propagator-denominator-mu} summarizes all the results obtained for the pole analysis.

\begin{itemize}
    \item In the case $\mu>0$:
    \begin{enumerate}
        \item $0<\kappa<\frac{2}{3\sqrt{3}}\frac{1}{\sqrt{\mu}}$: $f_{+}$ has two positive roots $m_{1}$ and $m_{2}$, whereas $f_{-}$ has only one root $m_{3}$,  satisfying the inequalities $ \kappa < m_{1} < \frac{1}{\sqrt{3\mu}} < m_{2} < \frac{1}{\sqrt{\mu}} < m_{3} < \frac{2}{\sqrt{3\mu}} $.
        \item $ \kappa=\frac{2}{3\sqrt{3}}\frac{1}{\sqrt{\mu}}$: $f_{+}$ has a double root $m_{1}=m_{2}=\frac{1}{\sqrt{3\mu}}$, and $f_{-}$ has one root $m_{3} = \frac{2}{\sqrt{3\mu}} $.
        \item $ \kappa > \frac{2}{3\sqrt{3}}\frac{1}{\sqrt{\mu}}$: $f_{+}$ has no positive root, whereas $f_{-}$ has only one positive root $m_{3} > \frac{2}{\sqrt{3\mu}} $.
    \end{enumerate}
\end{itemize}
\begin{itemize}
    \item In the case $\mu\le 0$:
    \begin{enumerate}
        \item $\kappa > 0$: $f_{+}$ has one positive root $m_{*} \le \kappa$ and $f_{-}$ has no positive root.
    \end{enumerate}
\end{itemize}

\begin{figure}[H]
    \centering
    \begin{subfigure}[t]{0.50\textwidth}
        \centering
        \scalebox{0.8}{
            \begin{tikzpicture}[xscale= 2.5, yscale=2.5]
                \coordinate (y-axis) at (0,1);
    
                \pgfmathsetmacro{\muval}{0.7};
    
                \pgfmathsetmacro{\k}{0.7 * 2/sqrt(27*\muval)};
    
                \draw[-{Stealth[scale=1.3]}] ({-1.2/sqrt(\muval)}, 0) -- ({1.7/sqrt(\muval)}, 0)
                node[below right] {$t$};
                
                \draw[-{Stealth[scale=1.3]}] (0, {-2.3*(2/sqrt(27*\muval))}) -- (0, {2.3*(2/sqrt(27*\muval)});
    
                
                \draw[] ({1/sqrt(\muval)}, 0.03) -- ({1/sqrt(\muval)}, -0.03) node[xshift=-2, yshift=-10] {$\tfrac{1}{\sqrt{\mu}}$};
                
                \draw[] ({-1/sqrt(\muval)}, 0.03) -- ({-1/sqrt(\muval)}, -0.03) node[xshift=-6, yshift=-10] {$-\tfrac{1}{\sqrt{\mu}}$};
    
                \draw[] ({1/sqrt(3*\muval)}, -0.03) -- ({1/sqrt(3*\muval)}, 0.03)
                node[above] {$\tfrac{1}{\sqrt{3\mu}}$};
    
                \draw[] ({2/sqrt(3*\muval)}, 0.03) -- ({2/sqrt(3*\muval)}, -0.03)
                node[xshift=6, yshift=-10] {$\tfrac{2}{\sqrt{3\mu}}$};
    
                \draw[] (\k, 0.03) -- (\k, -0.03)
                node[below] {$\kappa$};
                
                \draw[] (0.03, \k) -- (-0.03, \k)
                node[left] {$\kappa$};
    
                \draw[] (0.03, -\k) -- (-0.03, -\k)
                node[left=0] {$-\kappa$};

                \draw[] (-0.03, {2/sqrt(27*\muval)}) -- (0.03, {2/sqrt(27*\muval)});
                
                \begin{scope}[domain={-1.0/sqrt(\muval):1.2/sqrt(\muval)}]
                    \draw[line width=0.9, blue, smooth]
                    plot(\x, {\muval*\x^3 - \x + \k})
                    node[below right] {$f_{+}(t)$};
    
                    \draw[line width=0.7, gray, smooth, domain={-0.7/sqrt(\muval):1.0/sqrt(\muval)}]
                    plot(\x, {-\x + \k})
                    node[right, black] {$f_{+}(t),\;\; \mu=0$};
                    
                    \draw[line width=0.7, cyan, smooth, domain={-0.5/sqrt(\muval):0.8/sqrt(\muval)}]
                    plot(\x, {-\muval*\x^3 - \x + \k})
                    node[right] {$f_{+}(t), \;\; \mu<0$};
                    
                    \draw[line width=0.6, gray!80, smooth]
                    plot(\x, {\muval*\x^3 - \x})
                    node[black, below right] {$f_{0}(t)$};
    
                    \draw[line width=0.9, red, smooth]
                    plot(\x, {\muval*\x^3 - \x - \k})
                    node[below right] {$f_{-}(t)$};
                \end{scope}
    
                \draw[red!50, dashed]
                ({1/sqrt(3*\muval)}, 0) -- ({1/sqrt(3*\muval)}, {-2/sqrt(27*\muval) - \k});
    
                \draw[blue!50, dashed]
                ({-1/sqrt(3*\muval)}, 0) -- ({-1/sqrt(3*\muval)}, {2/sqrt(27*\muval) + \k});
    
                \draw[dotted]
                ({-1/sqrt(3*\muval)}, {2/sqrt(27*\muval)}) -- (0, {2/sqrt(27*\muval)})
                node[right, yshift=-1] {$\frac{2}{3\sqrt{3}}\;\frac{1}{\sqrt{\mu}}$};
    
            \end{tikzpicture}
        }
        \caption{Case $\mu > 0$, with $0 < \kappa < \frac{2}{3\sqrt{3}}\frac{1}{\sqrt{\mu}}$.}
        \label{fig:propagator-denominator-mu-positive}
    \end{subfigure}%
    \hfill
    \begin{subfigure}[t]{0.45\textwidth}
        \centering
        \scalebox{0.8}{
            \begin{tikzpicture}[xscale= 2.5, yscale=2.5]
                \coordinate (y-axis) at (0,1);
    
                \pgfmathsetmacro{\muval}{0.7};
    
                \pgfmathsetmacro{\k}{0.7 * 2/sqrt(27*\muval)};
    
                \draw[-{Stealth[scale=1.3]}] ({-0.6/sqrt(\muval)}, 0) -- ({0.8/sqrt(\muval)}, 0)
                node[below right] {$t$};
                
                \draw[-{Stealth[scale=1.3]}] (0, {-2.3*(2/sqrt(27*\muval))}) -- (0, {2.3*(2/sqrt(27*\muval)});
    
                
                
    
    
    
                \draw[] (\k, -0.03) -- (\k, 0.03)
                node[above] {$\kappa$};
                
                \draw[] (-\k, 0.03) -- (-\k, -0.03)
                node[below, xshift=-5] {$-\kappa$};
                
                \draw[] (-0.03, \k) -- (0.03, \k)
                node[right] {$\kappa$};
    
                \draw[] (0.03, -\k) -- (-0.03, -\k)
                node[left] {$-\kappa$};
    
                
                \begin{scope}[domain={-0.55/sqrt(\muval):0.58/sqrt(\muval)}]
                    
                    \draw[line width=0.9, blue, smooth]
                    plot(\x, {-\muval*\x^3 - \x + \k})
                    node[right] {$f_{+}(t)$};
    
                    \draw[line width=0.9, gray, smooth]
                    plot(\x, {-\muval*\x^3 - \x })
                    node[right, black] {$f_{0}(t)$};
    
                    \draw[line width=0.9, red, smooth]
                    plot(\x, {-\muval*\x^3 - \x - \k})
                    node[right] {$f_{-}(t)$};
                    
    
                \end{scope}

    
            \end{tikzpicture}
        }
        \caption{Case $\mu < 0$, with $\kappa > 0$.}
        \label{fig:propagator-denominator-mu-negative}
    \end{subfigure}
    \caption{Representative graphs of the cubic functions $f_\pm(t)=\mu t^{3}-t\pm\kappa$ and
    $f_{0}(t)=t(\mu t^{2}-1)$. (a): where
    $f_{+}$ has two positive real roots $(m_1,m_2)$ and $f_{-}$ has one $(m_3)$. (b): where a single positive real root is present.}
    \label{fig:propagator-denominator-mu}
\end{figure}

In the three-real-pole regime
$
0<\kappa^2\mu<\frac{4}{27},$
the roots satisfy
\begin{equation}
1-\mu m_1^2>0,\qquad 1-\mu m_2^2<0,\qquad 1-\mu m_3^2<0.
\label{eq:1-mum2-signs}
\end{equation}

At a massive pole $p^2=m^2$ we define $\sigma=\operatorname{sgn}\!\bigl[\kappa\,(1-\mu m^2)\bigr]$
so that $\sigma=+$ means the pole belongs to $D_+(p)=0$ and is saturated by $\|J^{(+)}\|^2$,
whereas $\sigma=-$ means it belongs to $D_-(p)=0$ and is saturated by $\|J^{(-)}\|^2$.
In the three-real-pole regime with \eqref{eq:1-mum2-signs}, one gets
\begin{equation}
\sigma_1=\sigma_2=\operatorname{sgn}(\kappa),
\qquad
\sigma_3=-\operatorname{sgn}(\kappa).
\label{eq:sigma123}
\end{equation}
For $\mu<0$ the unique massive pole always satisfies $1-\mu m_*^2>0$, hence
\begin{equation}
\sigma_*=\operatorname{sgn}(\kappa).
\label{eq:sigma-star}
\end{equation}

The sign of the residue at $p^2=m^2$ is controlled by $\operatorname{sgn}\!\left(1-3\mu m^2\right).$
Thus, in the three-pole regime one finds
\begin{equation}
\operatorname{sgn}\!\bigl(\operatorname{Res}\mathcal A\bigr)\Big|_{m_1}=+,
\qquad
\operatorname{sgn}\!\bigl(\operatorname{Res}\mathcal A\bigr)\Big|_{m_2,m_3}=-,
\end{equation}
whereas for $\mu<0$ the unique massive pole has $1-3\mu m_*^2>0$ and therefore positive residue. The derivation of (\ref{eq:residue_general_kappa_sign}) assumes that the physical pole is simple, i.e.
$\partial_{0}D(p_\ast)\neq 0$, or equivalently $G'(\lambda_\ast)\neq 0$.
At the critical point $\kappa^{2}\mu = 4/27$ the massive root becomes degenerate,
satisfying simultaneously
\begin{equation}
D_\sigma(p_\ast)=0,
\qquad
\partial_{0}D_\sigma(p_\ast)=0,
\end{equation}
which corresponds to a double zero of $G(\lambda)$ at
$\lambda_\ast = p_\ast^{2} = 1/(3\mu)$.
As a consequence, the linear approximation used in (\ref{eq:residue_general_kappa_sign}) is no longer valid
and the transverse contribution of propagator develops a second--order pole in $p^{0}$. Indeed, as shown in \cite{accioly}, double poles in the propagator rendering the theory non--unitary in this regime. For completeness, table \ref{tab2} summarize the pole structure and the corresponding residue signs obtained in the present analysis.

\begin{table}[H]
\centering
\caption{Poles structure in the physical region $p^2>0$.}
\label{tab2}
\footnotesize
\renewcommand{\arraystretch}{1.08}
\setlength{\tabcolsep}{3.5pt}

\begin{tabularx}{\linewidth}{c c Y c}
\hline
Condition 
& Poles in $p^2>0$ 
& Sector (vanishing $D_\pm$) 
& $\operatorname{sgn}(\mathrm{Res}\,\mathcal A)$\\
\hline

\rowcolor{lightgray}\multicolumn{4}{c}{$\mu>0$}\\
\hline

$0<\kappa^2\mu<\dfrac{4}{27}$
& $m_1,m_2,m_3$
& $\begin{array}{l}
m_{1,2}:\ \sigma=\operatorname{sgn}(\kappa)\\
m_{3}:\ \sigma=-\operatorname{sgn}(\kappa)
\end{array}$
& $\begin{array}{l}
m_{1}:\ +\ (1-3\mu m_1^2>0)\\
m_{2}:\ -\ (1-3\mu m_2^2<0)\\
m_{3}:\ -\ (1-3\mu m_3^2<0)
\end{array}$\\
\hline

$\kappa^2\mu=\dfrac{4}{27}$
& $m_1=m_2=\dfrac{1}{\sqrt{3\mu}},\ m_3$
& same criterion as above
& $m_{1,2}:\text{undefined}\ ;\quad m_3:-$\\
\hline

$\kappa^2\mu>\dfrac{4}{27}$
& $m_3$
& $\sigma=-\operatorname{sgn}(\kappa)$
& $-$\\
\hline

\rowcolor{lightgray}\multicolumn{4}{c}{$\mu<0$}\\
\hline

$\kappa^2\mu<0$
& $m_*$
& $\sigma=\, \operatorname{sgn}(\kappa)$
& $+$\\
\hline
\end{tabularx}
\end{table}

The pole structure summarized in table \ref{tab2} is in agreement with the results previously obtained for Lee--Wick--Chern--Simons-type electrodynamics. In particular, the higher-derivative sector produces additional massive poles whose residue signs cannot all be made positive simultaneously in the $\mu>0$ sector, reproducing the ghostlike behavior identified in earlier analyses of planar Lee--Wick and Maxwell--Chern--Simons extensions~\cite{accioly, grinstein2008,carone2009,borges2022mcs, borges2022lw}. In the present formulation, the same conclusion follows directly from the spectral decomposition in the projector basis $(P_1,P_2,P_3)$: the sign of the residue is controlled by the factor $1-3\mu m_i^2$, making explicit which of the transverse contributions carries negative norm.

Thus, although a given pole of the transverse propagator may satisfy
the positivity condition of the residue in a restricted region of the
parameter space, tree-level unitarity requires all propagating modes to have
positive residues simultaneously. A detailed analysis of the dispersion
relation shows that this condition cannot be fulfilled in the
Maxwell--Lee--Wick--Chern--Simons model.

Restricting ourselves to the $\mu>0$ sector, whenever the spectrum contains three
real poles, two of them necessarily lie in the region
$\lambda>1/3\mu$ and carries a negative residue. Likewise, in the
$\mu>0$ regime with a single real pole, the latter is also located at $\lambda>1/3\mu$, again
corresponding to a ghost excitation. The sign of $\kappa$ does not affect this
conclusion, as it merely selects which transverse contribution saturates the
pole, without altering the location of the roots or the sign structure of the
residues. Therefore, no choice of parameters renders all propagating modes
unitary, and the Maxwell--Lee--Wick--Chern--Simons model is intrinsically
nonunitary at tree level.

Finally, let us comment on the $\mu<0$ sector. In this case, the spectrum
also contains complex poles, similarly to the $\kappa ^2 \mu>4/27$ regime discussed
above. However, the massive real pole that survives in the physical region
$p^2>0$ carries a positive residue, independently of the sign of $\kappa$.
This feature distinguishes the $\mu<0$ sector from the corresponding
$\mu>0$ configurations with a single real pole, where the latter is always
associated with a ghost excitation. Nevertheless, the presence of complex
poles and the lack of a parameter choice yielding a fully healthy spectrum
prevent the restoration of tree-level unitarity in the model.

\subsubsection{Causality}

For parameter values such that all poles $\lambda$ are
real and positive ($\kappa^2\mu<4/27$), the associated modes
have relativistic dispersion relations and subluminal group velocities,
preserving microscopic causality. When
$\kappa^2\mu>4/27$, two of the poles become complex conjugates;
in this situation the coordinate-space propagator exhibits
damped oscillations, corresponding to unstable resonances.
Following the Lee--Wick contour prescription, one chooses the
integration contour in $p^0$ so that such modes decay in the
future, preserving the unitarity of the $S$-matrix and macroscopic
causality, at the cost of small local violations of
microcausality, suppressed by the Lee--Wick mass scale.

\subsection{Analysis for the Maxwell--Deser--Jackiw model}

Using (\ref{abgMDJ}), the poles are
\begin{equation}
p^4=0\Rightarrow p^2=0,\qquad p^2=m^2.
\end{equation}
The pole at $p^2=0 \Rightarrow p^0=\pm\|\vec{p}\|$ is purely a gauge pole and does not contribute to
observables when the propagator is saturated with a conserved
current. The physically relevant pole is therefore
\begin{equation}
p^2 = m^2\Rightarrow p^0=\pm E_*(\vec{p})=\pm \sqrt{\vec p^{\,2}+m^2},
\end{equation}
which describes a single massive mode with mass $m>0$ and relativistic energy $E_*(\vec{p})$.

\subsubsection{Saturated amplitude and residues}

Using the projector decomposition introduced in (\ref{eq:A_general_P}),
the saturated amplitude can be written as

\begin{equation}
\mathcal A(p^0,\vec p)
=
\frac{1}{\big((p^0)^2-\vec p^{\,2}\big)}\left[\frac{\|J^{(+)}\|^{2}}{\left(1-\frac{\sqrt{(p^0)^2-\vec p^{\,2}}}{m}\right)}
+
\frac{\|J^{(-)}\|^{2}}{\left(1+\frac{\sqrt{(p^0)^2-\vec p^{\,2}}}{m}\right)}\right].
\label{AJpm}
\end{equation}

Near the pole we have

\begin{equation}
    \mathcal A(p^{0},\vec p)\Big|_{\,p^{0}=E_*(\vec p)}
\simeq-\frac{\|J^{(+)}\|^{2}}{E_*(\vec p)}\,
\frac{1}{p^{0}-E_*(\vec p)}+
\frac{\|J^{(-)}\|^{2}}{2m^2}.
\end{equation}
This shows that only the positive term propagates, whereas the
negative is projected out on the mass shell. The residue at
$p^o=E_*(\vec p)=\sqrt{\vec p^{\,2}+m^{2}}$ is

\begin{equation}
\operatorname*{Res}_{\,p^{0}=E_*(\vec p)} \mathcal A(p^{0},\vec p)
=-\frac{\|J^{(+)}\|^{2}}{E_*(\vec p)}.
\end{equation}
Since $E_*(\vec{p})>0$ and $\|J^{(+)}\|^{2}\geq 0$, the residue is negative. Therefore, the massive mode associated with this pole is a ghost and the model is not unitary at the tree level.

\subsubsection{Causality}

Since the only physical pole is simple and real, with
\begin{equation}
(p^0)^2-\vec{p}^{\,2} = m^2\Rightarrow p^0=\pm E_*(\vec{p})\Rightarrow
E_*(\vec{p}) = \sqrt{\vec p^{\,2} + m^2}.
\end{equation}
For the positive–energy branch, the group velocity satisfies

\begin{equation}
v_{g}
=
\left|\frac{\partial E_*(\vec p)}{\partial \vec p}\right|
=
\frac{|\vec p|}{\sqrt{\vec p^{\,2}+m^{2}}}
< 1,
\end{equation}
in
Lorentz units. So there is no superluminal propagation. Moreover, the
absence of complex poles indicates that there are no Lee--Wick type
modes; the operator $S_{\mu\nu}$ merely selects a single propagating mode without violating microcausality. Therefore, the Maxwell–Deser–Jackiw model provides an example in which causality and tree-level unitarity do not coexist: the theory exhibits a single causal massive excitation, but the associated negative residue signals a ghost mode, rendering the model nonunitary. This result is consistent with previous studies of higher-derivative Chern--Simons electrodynamics and related planar gauge models, in which the presence of a real massive pole does not by itself guarantee tree-level unitarity~\cite{boldo,deserjackiw,avila2020,lisboasantos2023}. In those approaches, the physical acceptability of the excitation is determined by the sign of the residue of the saturated propagator or, equivalently, by the positivity of the corresponding norm in the physical subspace. The projector formulation used here reproduces this conclusion in a particularly transparent way: the massive Maxwell--Deser--Jackiw excitation has a relativistic dispersion relation and therefore satisfies the causality criterion, but its negative residue identifies it as a ghost.

\subsection{Analysis for the Nonlocal
Maxwell--Chern--Simons Model}
\label{subsec:6.4}

We now investigate the nonlocal Maxwell--Chern--Simons
model introduced in Section~\ref{subsec:5.4},
using the projection-operator formalism developed
in the preceding sections.
In particular, we start from the coefficients
of the wave operator obtained in~\eqref{abgn}. For a conserved external current,
the longitudinal contribution proportional to
$P_1$ does not enter the saturated propagator.
Consequently, the relevant transverse denominator is given by~\eqref{eq:nonlocal_denominator}. Thus, the nonzero massive poles are determined by
\begin{equation}
p^2e^{2p^2/M^2}
-\kappa^2f(-p^2)^2=0.
\label{eq:analysis_nonlocal_poles}
\end{equation}
Assuming $M^2>0$, a real and nonzero
Chern--Simons parameter $\kappa$, and a function
$f$ with real Taylor coefficients that is analytic
in the relevant domain,~\eqref{eq:analysis_nonlocal_poles}
provides the starting point for analyzing the pole
structure and the corresponding physical properties
of the model.

\subsubsection{Saturated amplitude and residues}

For an external conserved current, the longitudinal
part of the inverse wave operator vanishes upon saturation. In this way~\eqref{eq:A_general_P} of
the general analysis gives
\begin{equation}
\begin{aligned}
\mathcal A(p^0,\vec p)
={}&
\frac{\|J^{(+)}\|^2}{
\bigl[(p^0)^2-\vec p^{\,2}\bigr]
e^{[(p^0)^2-\vec p^{\,2}]/M^2}
-\kappa f\!\left(\vec p^{\,2}-(p^0)^2\right)
\sqrt{(p^0)^2-\vec p^{\,2}}
}
\\[2mm]
&+
\frac{\|J^{(-)}\|^2}{
\bigl[(p^0)^2-\vec p^{\,2}\bigr]
e^{[(p^0)^2-\vec p^{\,2}]/M^2}
+\kappa f\!\left(\vec p^{\,2}-(p^0)^2\right)
\sqrt{(p^0)^2-\vec p^{\,2}}
}.
\end{aligned}
\label{eq:nl63_amp}
\end{equation}
The two scalar denominators are denoted by
\begin{equation}
\begin{aligned}
D_+(p)&=\beta(p)-i\gamma(p)\sqrt{p^2}
=-p^2e^{p^2/M^2}+\kappa f(-p^2)\sqrt{p^2},\\
D_-(p)&=\beta(p)+i\gamma(p)\sqrt{p^2}
=-p^2e^{p^2/M^2}-\kappa f(-p^2)\sqrt{p^2}.
\end{aligned}
\label{eq:nl63_Dpm}
\end{equation}
Now, we define the product of the two
projector denominators by
\begin{equation}
\begin{aligned}
D(p^0,\vec p)
&=D_+(p)D_-(p)
 =\beta(p)^2+\gamma(p)^2p^2\\
&=\bigl[(p^0)^2-\vec p^{\,2}\bigr]
G\!\left((p^0)^2-\vec p^{\,2}\right),
\end{aligned}
\label{eq:nl63_D}
\end{equation}
where the scalar function $G(p^2)$ is explicitly given by
\begin{equation}
G(p^2)=p^2e^{2p^2/M^2}-\kappa^2f(-p^2)^2.
\label{eq:nl63_G}
\end{equation}
The nonzero real roots of $G(p^2)$
determine the candidate massive poles, whereas the explicit factor
$p^2$ must be treated separately when analyzing the lightlike sector.
Let $m_i>0$ denote the mass of the $i$th massive mode, so that
$p^2=m_i^2$ at the corresponding pole.
A root contributes to
the $P_2$ or $P_3$ term according to which denominator in~\eqref{eq:nl63_amp} vanishes. Equivalently,
\begin{equation}
m_i\,e^{m_i^2/M^2}=\pm\kappa f(-m_i^2).
\label{eq:nl63_projector_poles}
\end{equation}
Squaring~\eqref{eq:nl63_projector_poles} does not distinguish between the two projector contributions.
For $m_i>0$, the two denominators cannot vanish simultaneously. For each simple positive mass-shell root $p^2=m_i^2$ of $G(p^2)=0$, the corresponding
positive-energy solution is
\begin{equation}
E_*(\vec p)=\sqrt{\vec p^{\,2}+m_i^2},\qquad
p^0=E_*(\vec p).
\label{eq:nl63_energy}
\end{equation}
The saturated amplitude can also be written over the common denominator
$D(p^0,\vec p)=p^2G(p^2)$:
\begin{equation}
\begin{aligned}
\mathcal A(p^0,\vec p)
={}&\frac{\left[p^2e^{p^2/M^2}+\kappa f(-p^2)\sqrt{p^2}\right]
\|J^{(+)}\|^2}{p^2G(p^2)}\\[2mm]
&+\frac{\left[p^2e^{p^2/M^2}-\kappa f(-p^2)\sqrt{p^2}\right]
\|J^{(-)}\|^2}{p^2G(p^2)}.
\end{aligned}
\label{eq:nl63_common}
\end{equation}
At a $P_2$ pole the numerator multiplying $\|J^{(+)}\|^2$
becomes $2m_i^2 e^{m_i^2/M^2}$, while the other term vanishes;
at a $P_3$ pole the converse holds. Consequently the residue at $p^0=E_*(\vec p)$ is
\begin{equation}
\begin{aligned}
\underset{p^0=E_*(\vec p)}{\operatorname{Res}}
\mathcal A(p^0,\vec p)
&=\frac{e^{m_i^2/M^2}}{G'(m_i^2)E_*(\vec p)}
\|J^{(\sigma)}\|^2,\\[2mm]
\sigma&=\begin{cases}
+, & \kappa f(-m_i^2)>0,\\
-, & \kappa f(-m_i^2)<0.
\end{cases}
\end{aligned}
\label{eq:nl63_residue}
\end{equation}
where
\begin{equation}
G'(m_i^2)=e^{2m_i^2/M^2}\left(1+\frac{2m_i^2}{M^2}\right)
+2\kappa^2f(-m_i^2)f'(-m_i^2).
\label{eq:nl63_derivative}
\end{equation}
Here $f'$ denotes differentiation with respect to the argument of $f$.
Since $E_*(\vec p)>0$ and $e^{m_i^2/M^2}>0$, the sign of the
residue is determined by
\begin{equation}
\operatorname{sgn}\!\left(\operatorname{Res}\mathcal A\right)
=\operatorname{sgn}\!\left[\frac{1}{G'(m_i^2)}\right].
\label{eq:nl63_residue_sign}
\end{equation}
The sign of the residue at a positive-energy simple massive pole is
therefore positive precisely when
\begin{equation}
e^{2m_i^2/M^2}\left(1+\frac{2m_i^2}{M^2}\right)
+2\kappa^2f(-m_i^2)f'(-m_i^2)>0.
\label{eq:nl63_unitarity}
\end{equation}
Using~\eqref{eq:nl63_projector_poles}, this can also be expressed as
\begin{equation}
1+\frac{2m_i^2}{M^2}
+2m_i^2\frac{f'(-m_i^2)}{f(-m_i^2)}>0,
\label{eq:nl63_log_condition}
\end{equation}
where $f(-m_i^2)\ne0$, as is automatic for the massive roots under
our assumptions.

\paragraph{Example I: $f(\Box)=1$.}
In this case, the equation for the real massive poles reduces to
\begin{equation}
p^2e^{2p^2/M^2}-\kappa^2=0.
\label{eq:nl63_ex1}
\end{equation}
In analogy with the analysis of the polynomial models, we introduce
$t=p^2/M^2>0$ and the dimensionless parameter
$\beta=\kappa^2/M^2>0$. The pole equation then reads
\begin{equation}
g(t)-\beta=0,\qquad
g(t)=te^{2t},\qquad t>0.
\label{eq:nl63_ex1_function}
\end{equation}
Its first derivative is
\begin{equation}
g'(t)=e^{2t}(1+2t)>0,\qquad t>0.
\label{eq:nl63_ex1_derivative}
\end{equation}
Since $g(0)=0$, $g(t)$ is strictly increasing for $t>0$,
and $g(t)\to\infty$ as $t\to\infty$, every $\beta>0$
corresponds to exactly one simple positive real root $t_1$.
The solution is expressed through the principal real branch
of the Lambert function:
\begin{equation}
\begin{aligned}
t_1&=\frac12 W_0(2\beta),&
m_1^2&=M^2t_1=\frac{M^2}{2}W_0(2\beta).
\end{aligned}
\label{eq:nl63_ex1_lambert}
\end{equation}
Figure~\ref{fig:nl63_spectrum} illustrates the unique
intersection of $g(t)$ with each positive value of $\beta$.
At the positive real massive pole, the projector label obeys
\begin{equation}
\sigma_1=
\begin{cases}
+, & \kappa>0,\\
-, & \kappa<0,
\end{cases}
\label{eq:nl63_ex1_projector}
\end{equation}
because $f(-m_1^2)=1$. Its residue sign follows from
\begin{equation}
G'(m_1^2)=e^{2t_1}(1+2t_1)>0,\qquad
\operatorname{sgn}(\operatorname{Res}\mathcal A)\big|_{m_1^2}=+.
\label{eq:nl63_ex1_signs}
\end{equation}

\begin{figure}[htbp]
\centering

\begin{tikzpicture}
\begin{axis}[
    width=0.83\linewidth,
    height=5.8cm,
    xlabel={$t=p^2/M^2$},
    ylabel={$g(t)$},
    xmin=0, xmax=1.25,
    ymin=0, ymax=1.25,
    domain=0:1.25,
    samples=180,
    axis line style={black},
    tick style={black},
    tick label style={black},
    label style={black},
    legend style={
        at={(0.98,0.98)},
        anchor=north east,
        font=\small,
        draw=black,
        fill=white
    },
    grid=major,
    grid style={gray!25}
]

\addplot[
    blue,
    very thick,
    solid
] {x*exp(2*x)};
\addlegendentry{$g(t)=te^{2t}$}

\addplot[
    red,
    thick,
    dashed
] {0.10};
\addlegendentry{$\beta=0.10$}

\addplot[
    green!60!black,
    thick,
    densely dotted
] {0.30};
\addlegendentry{$\beta=0.30$}

\addplot[
    magenta,
    thick,
    dashdotted
] {0.70};
\addlegendentry{$\beta=0.70$}

\end{axis}
\end{tikzpicture}

\caption{Graphical representation of
$g(t)=te^{2t}$ for different values
of $\beta$.}

\label{fig:nonlocal_spectrum}

\end{figure}

For every $\beta>0$, the unique simple massive pole has a
positive residue:
\begin{equation}
\operatorname{sgn}(\operatorname{Res}\mathcal A)\big|_{m_1^2}=+.
\label{eq:nl63_ex1_two_signs}
\end{equation}
In particular, the positive-energy residue is
\begin{equation}
\operatorname*{Res}_{p^0=E_*(\vec p)}\mathcal A
=\frac{e^{-t_1}}{(1+2t_1)E_*(\vec p)}
\|J^{(\sigma_1)}\|^2>0
\end{equation}
for a nonvanishing source projection.
There is no critical value of $\beta$ at which two positive
real poles merge. The unique positive real pole remains simple
for all $\beta>0$. This positive-residue result is restricted
to the real massive pole and does not by itself establish
full quantum unitarity or stability in the presence of
possible complex poles.

\paragraph{Example II: $f(\Box)=e^{-\Box/M^2}$.}
For this choice, the two nonlocal factors coincide:
$f(-p^2)=e^{p^2/M^2}$. The pole equation factorizes as
\begin{equation}
G(p^2)=e^{2p^2/M^2}(p^2-\kappa^2)=0.
\label{eq:nl63_ex2}
\end{equation}
The exponential has no finite zeros, so there is precisely
one finite massive pole for $\kappa\ne0$:
\begin{equation}
m_*^2=\kappa^2,\qquad
E_*(\vec p)=\sqrt{\vec p^{\,2}+\kappa^2}.
\label{eq:nl63_ex2_pole}
\end{equation}
The corresponding projector is selected by
\begin{equation}
\sigma_*=
\begin{cases}
+, & \kappa>0,\\
-, & \kappa<0,
\end{cases}
\label{eq:nl63_ex2_projector}
\end{equation}
because the exponential factor is strictly positive.
Differentiation of~\eqref{eq:nl63_ex2} yields
\begin{equation}
G'(m_*^2)=e^{2\kappa^2/M^2}>0.
\label{eq:nl63_ex2_derivative}
\end{equation}
Accordingly, the positive-energy residue is
\begin{equation}
\operatorname*{Res}_{p^0=E_*(\vec p)}
\mathcal A(p^0,\vec p)
=\frac{e^{-\kappa^2/M^2}}{E_*(\vec p)}
\|J^{(\sigma_*)}\|^2>0
\label{eq:nl63_ex2_res}
\end{equation}
for a nonvanishing source projection.
Thus the massive pole structure is the same as that of local
Maxwell--Chern--Simons theory, while the residue is multiplied
by the positive factor $e^{-\kappa^2/M^2}$.
As in Example I, no additional positive real massive pole
is introduced. The massless sector must still be considered
separately because the projector decomposition is singular
at $p^2=0$. Furthermore, positivity of the single massive
residue is a tree-level statement; it does not, by itself,
prove microcausality or full quantum unitarity in a nonlocal
theory. Table~\ref{tab:nl63_comparison} summarizes the results
obtained for the two nonlocal models considered.

\begin{table}[htbp]
\centering
\caption{Comparison of the positive real massive spectra
for the two nonlocal form factors.}
\label{tab:nl63_comparison}

\vspace{0.2cm}

\begingroup
\footnotesize
\renewcommand{\arraystretch}{1.5}
\setlength{\tabcolsep}{4pt}

\begin{tabular}{@{}
>{\raggedright\arraybackslash}p{0.22\linewidth}
>{\raggedright\arraybackslash}p{0.43\linewidth}
>{\centering\arraybackslash}p{0.18\linewidth}
>{\centering\arraybackslash}p{0.11\linewidth}
@{}}

\toprule
\textbf{Function} &
\textbf{Real massive poles} &
\textbf{$\sigma$ (by pole)} &
\textbf{Residue signs} \\

\midrule

$f(\Box)=1$ &
One simple pole,
$\displaystyle
m_1^2=\frac{M^2}{2}
W_0\!\left(\frac{2\kappa^2}{M^2}\right)
$ &
$\operatorname{sgn}(\kappa)$ &
$(+)$ \\

\addlinespace[5pt]

$f(\Box)=e^{-\Box/M^2}$ &
One simple pole,
$\displaystyle m_*^2=\kappa^2$ &
$\operatorname{sgn}(\kappa)$ &
$(+)$ \\

\bottomrule

\end{tabular}
\endgroup
\end{table}

\paragraph{General function.}
For an arbitrary real analytic $f$, the pole positions satisfy
\begin{equation}
m_i^2 e^{2m_i^2/M^2}=\kappa^2f(-m_i^2)^2,
\qquad m_i^2>0.
\label{eq:nl63_general}
\end{equation}
The projector contribution is selected by
$\sigma_i=\operatorname{sgn}[\kappa f(-m_i^2)]$, and the residue
sign is determined by $G'(m_i^2)$ as in~\eqref{eq:nl63_residue_sign}. Without further restrictions
on $f$, neither the number of positive real poles nor the
complete complex spectrum can be classified universally.

\subsubsection{Causality}\label{subsec6.4.2}
For each real positive root $x_i$, the pole position is $p^2=m_i^2=M^2x_i$.
Therefore, the dispersion relationship is
\begin{equation}
p^0=\pm E_*(\vec p),\qquad E_*(\vec p)=\sqrt{\vec p^{\,2}+m_i^2},\qquad m_i^2=M^2x_i.
\label{eq:nl63_dispersion}
\end{equation}
For the positive-energy branch, differentiating the mass-shell~\eqref{eq:nl63_dispersion} implicitly, and assuming a simple root, gives
\begin{equation}
\frac{\partial E_*(\vec p)}{\partial p_j}=\frac{p_j}{E_*(\vec p)},\qquad
v_g=\left|\frac{\partial E_*(\vec p)}{\partial\vec p}\right|=\frac{|\vec p|}{E_*(\vec p)}<1.
\label{eq:nl63_vg}
\end{equation}
Therefore all simple positive real mass poles have subluminal group
velocity. For real spacelike $p^2<0$, the quantity
\begin{equation}
p^2e^{2p^2/M^2}-\kappa^2f(-p^2)^2
\label{eq:nl63_spacelike}
\end{equation}
is strictly negative, provided $f(-p^2)$ is real. Accordingly, there
are no real tachyonic roots under these assumptions. For a function
with real Taylor coefficients, nonreal zeros of the mass-shell
equation occur in complex-conjugate pairs. Their interpretation
requires a prescription for the energy contour and boundary
conditions; neither the absence of tachyonic roots nor the
subluminal group velocity proves stability or full unitarity.
Finally, $p^2=0$ must be examined separately. The transverse and
longitudinal operators, as well as the operators $P_2$ and $P_3$, are
singular in their usual representations on the light cone. Although~\eqref{eq:nl63_D} contains an explicit factor $p^2$, the
massive factor evaluates to $-\kappa^2$ at $p^2=0$ because $f(0)=1$.
Saturation removes the longitudinal gauge contribution, but the
remaining antisymmetric term contains a $1/p^2$ factor and must be
studied with conserved sources in a regular lightlike limit.

It is worth noting that the local models previously
investigated can be recovered through suitable polynomial
truncations of the nonlocal functions. In particular,
the zeroth-order truncation of both functions reproduces
the Maxwell--Chern--Simons model, whereas the first-order
truncation of the exponential Maxwell function yields
the Maxwell--Lee--Wick--Chern--Simons model, with
$\mu_{\mathrm{LW}}=-M^{-2}$. Furthermore, retaining the
zeroth-order Maxwell term and the first-order
Chern--Simons function, with $c_0=0$ and
$c_1=-M^2/(\kappa m)$, recovers the
Maxwell--Deser--Jackiw model. Thus, the nonlocal
formulation encompasses the local theories examined
in this work as particular polynomial truncations,
with appropriate choices of coefficients.

\section{Conclusions and Outlook}\label{sec7}
In this article we developed a systematic construction of projection operators for
studying causality and unitarity in planar gauge theories with
higher-derivative and Chern--Simons terms. By expressing the momentum-space
wave operator in the usual $\{\omega,\theta,S\}$ basis with arbitrary
momentum-dependent coefficients and subsequently constructing a set of projection operators that decompose the relevant subspace into a direct sum of subspaces, we derived the propagator in the general-coefficient framework and obtained explicit conditions on its pole structure,
residues and analytic behavior. The resulting projector formalism was then applied to the study of the physical excitations,
ghostlike modes and tachyons. The proposed construction was shown to reproduce the expected features of the Maxwell--Lee--Wick--Chern--Simons
and Maxwell--Deser--Jackiw models, providing a unified perspective on
their causal and unitary properties.

The causality and unitarity analysis can be equivalently formulated in the projector basis $(P_1, P_2, P_3)$ or in the usual basis $(\omega, \theta, S)$, both leading to the same physical results while emphasizing different aspects of the theory. A distinctive feature of the present construction is that the operators
$(P_1,P_2,P_3)$ are genuine projection operators in the algebraic sense. This property does not hold for the usual
$\{\omega,\theta,S\}$ basis, since although $\omega$ and $\theta$ are projection
operators, $S$ is not a projector. Although conceptually distinct, the two formulations are mathematically equivalent and preserve all pole positions and residues, thus offering complementary perspectives on the same underlying physics.

Another interesting outcome of the present analysis of the Maxwell--Lee--Wick--Chern--Simons model is the appearance of the sector corresponding to negative values of the parameter $\mu$. This possibility emerges naturally within the proposed framework and can be investigated on the same way as the positive-$\mu$ sector. To the best of our knowledge, this case has received little attention in the existing literature and deserves a more detailed investigation.

A further significant application of the proposed
formalism concerns the nonlocal Maxwell--Chern--Simons
model, demonstrating that the new projection-operator
construction can be extended beyond local
higher-derivative theories to systems involving
nonpolynomial functions of the d'Alembertian
operator. In this context, the propagator was
obtained without modifying the underlying
projector algebra, and its pole structure,
tree-level unitarity, and causality properties
were investigated for representative choices
of the nonlocal functions. The analysis revealed
that the nonlocal modifications can substantially
influence the massive spectrum and the residues
of the saturated propagator, while preserving
the applicability of the proposed algebraic
framework. In particular, the examples considered
exhibit a single positive real massive pole
with a positive residue, providing a consistent
tree-level spectral interpretation within the
real massive sector.

Several natural extensions of this work deserve further study. In particular, the nonlocal
Maxwell--Chern--Simons formulation investigated
here opens the possibility of exploring more
general nonpolynomial functions and their
implications for the analytic structure of
the propagator, including complex poles and
UV behavior. Another way is the incorporation of interactions, either
perturbatively or through resummation techniques, in order to determine how
loop corrections modify the causal structure or shift the position of
Lee--Wick poles. It would also be interesting to apply the method to
gravitational models in $(2+1)$ dimensions, where higher-derivative and
topological terms play a central role and where projection operators provide a useful framework.

Finally, the generalized analytic strategy developed here is well suited to
examining whether causality and unitarity can coexist in more exotic
settings, such as parity-violating dualities, Lorentz-violating effective
actions, or finite-temperature backgrounds. We expect that the interplay
between higher-derivative terms and topological masses in these regimes
will continue to reveal rich structures, making planar gauge theories an
ideal laboratory for probing fundamental aspects of quantum field theory.


\appendix

\section{Supplementary Material}

The proofs of theorems \ref{theoremdirectsum}, \ref{theoremnucimg} and \ref{theoreminvO} can be found in \cite{lang,lima_algebra,axler2024}.\\

\begin{proof}\textbf{[Proposition \ref{proposition1}]}
If $v \in \text{Im}\,\omega \implies v = \omega(w) \implies S(v) = S(\omega(w)) = 0$, since $S\omega = 0$. 
Thus, we see that if $v \in \text{Im}\,\omega$, then $v \in \ker S$. 
In this way, the image of $\omega$ is invariant under $S$. 
It is easy to see that $\text{Im}\,\omega$ is also invariant under both $\omega$ and $\theta$. 
Indeed, note that if $v \in \text{Im}\,\omega$, then
\[
\omega(v) = v \in \text{Im}\,\omega,
\]
and
\[
\theta(v) = (I - \omega)(v) = I(v) - \omega(v) = v - \omega(v) = v - v = 0.
\]
Now, setting $W = \text{Im}\,\theta = \ker \omega$, we show that $S(W) \subseteq W$. 
Let $w \in W$ 
\[
w = \theta(v) \implies S(w) = S(\theta(v)) \implies S(w) = \theta(S(v)) \in W,
\]
where we used the fact that $S$ and $\theta$ commute. From now on, we will only consider the restrictions of $S$ and $\theta$ to the subspace $W$ and we use the notation $S_W = S|_W : W \to W$ and $\theta_W = \theta|_W = \mathbf{1}_W$, so that
\[
S_W^2 = -p^2 \mathbf{1}_W.
\]
Defining $Z : W \to W$ as $Z = a S_W + b \mathbf{1}_W$, we want to find $a,b$ such that $Z$ is a projector, that is
\begin{align*}
Z^2 &= Z \nonumber\\
(a S_W + b \mathbf{1}_W)^2
&= a S_W + b \mathbf{1}_W \nonumber\\
&= a^2 S_W^2 + 2ab S_W + b^2 \mathbf{1}_W \nonumber\\
&= \left(-a^2 p^2 + b^2\right) \mathbf{1}_W + 2ab S_W.
\end{align*}
thus,
\[
b = \frac{1}{2}, \quad a = \frac{i}{2\sqrt{p^2}}.
\]
Therefore,
\[
Z= \frac{1}{2} \mathbf{1}_W + \frac{i}{2\sqrt{p^2}} S_W,
\]
is a projector and, consequently, $W = \text{Im} \,Z \oplus \ker Z$. 
We can then write
\[
S_W
= -\left(Z - \frac{1}{2} 1_W \right) 2i\sqrt{p^2}.
\]
If $w \in \text{Im}\,Z$, then
\begin{align*}
S_W(w)
&= -\left(Z(w) - \frac{1}{2} 1_W(w) \right) 2i\sqrt{p^2} \nonumber\\
&= -\left(w - \frac{1}{2} w\right) 2i\sqrt{p^2} \nonumber\\
&= -i\sqrt{p^2}\, w \nonumber\\
&= S(w).
\end{align*}
Similarly, if $w \in \ker Z$, then
\[
S_W(w) = i\sqrt{p^2} w = S(w).
\]
Therefore, we see that vectors in $\text{Im}\, Z$ and $\ker Z$ are eigenvectors of $S$. 
Now, let us show that $\text{Im}\, Z$ and $\ker Z$ cannot be trivial subspaces! For this to be true, we must have $\dim(\text{Im}\, Z) = \dim(\ker Z) = 1$. 
Suppose initially that one of them is trivial and we will arrive at a contradiction.\\

\noindent \textbf{Case 1:} $\text{Im}\, Z = W$. 
Thus, $\forall \, w \in W $, we have
\[
S_W(w) = -i\sqrt{p^2} w \implies S_W = -i\sqrt{p^2} \mathbf{1}_W.
\]

\noindent \textbf{Case 2:} $\ker Z = W$. 
Thus, $\forall \, w \in W $, we have
\[
S_W(w) = i\sqrt{p^2} w \implies S_W = i\sqrt{p^2} \mathbf{1}_W.
\]
In general, $S_W = a \mathbf{1}_W$. Since $v = \omega(v) + \theta(v)$, it follows that
\begin{align*}
S(v) 
&= S(\omega(v)) + S(\theta(v)) = S(\theta(v)) \nonumber\\
&= S_W(\theta(v)) \nonumber\\
&= a\, \mathbf{1}_W(\theta(v)) \nonumber\\
&= a\, \theta(v),
\end{align*}
which gives us
\[
S = a \theta.
\]
Now let us make the following analysis,
\[
S^\mu_{\;\nu} = a \theta^\mu_{\;\nu}, \quad S_{\mu\nu} = a \theta_{\mu\nu}.
\]
Since $\theta$ is symmetric and $S_{\mu\nu}$ is skew-symmetric, we obtain
\[
S_{\mu\nu} = a \theta_{\nu\mu}= S_{\nu\mu} = -S_{\mu\nu},
\]
which implies $S_{\mu\nu} = \varepsilon_{\mu\nu \beta} \, p^{\beta} = 0$, and so $p = 0$. 
But this is absurd, since we suppose $p \neq 0$.
\end{proof}

\bibliographystyle{unsrtnat}
\bibliography{refs}

\end{document}